\documentclass[10pt,twocolumn,letterpaper]{article}

\usepackage[pagenumbers]{cvpr} 

\definecolor{cvprblue}{rgb}{0.21,0.49,0.74}
\usepackage[pagebackref,breaklinks,colorlinks,allcolors=cvprblue]{hyperref}
\usepackage{hyperref}
\usepackage{cleveref}
\usepackage{xcolor}
\def\paperID{*****} 

\title{A Heterogeneous Ensemble for Multi-Center COVID-19 Classification from Chest CT Scans}

\author{%
  \begin{tabular}[t]{@{}cc@{}}
    Aadit Nilay & Bhavesh Thapar \\[2pt]
    {\tt\small aadit@terpmail.umd.edu} & {\tt\small bthapar@terpmail.umd.edu} \\[6pt]
    Anant Agrawal & Mohammad Nayeem Teli \\[2pt]
    {\tt\small anant04@terpmail.umd.edu} & {\tt\small nayeem@umd.edu} \\[4pt]
    \multicolumn{2}{c}{University of Maryland, College Park}
  \end{tabular}
}

\begin{document}
\maketitle

\begin{abstract}
The COVID-19 pandemic exposed critical limitations in diagnostic workflows: RT-PCR tests suffer from slow turnaround times and high false-negative rates, while CT-based screening offers faster complementary diagnosis but requires expert radiological interpretation. Deploying automated CT analysis across multiple hospital centres introduces further challenges, as differences in scanner hardware, acquisition protocols, and patient populations cause substantial domain shift that degrades single-model performance. To address these challenges, we present a heterogeneous ensemble of nine models spanning three inference paradigms: (1)~a self-supervised DINOv2 Vision Transformer with slice-level sigmoid aggregation, (2)~a RadImageNet-pretrained DenseNet-121 with slice-level sigmoid averaging, and (3)~seven Gated Attention Multiple Instance Learning models using EfficientNet-B3, ConvNeXt-Tiny, and EfficientNetV2-S backbones with scan-level softmax classification. Ensemble diversity is further enhanced through random-seed variation and Stochastic Weight Averaging. We address severe overfitting, reducing the validation-to-training loss ratio from 35$\times$ to less than 3$\times$, through a combination of Focal Loss, embedding-level Mixup, and domain-aware augmentation. Model outputs are fused via score-weighted probability averaging and calibrated with per-source threshold optimization. The final ensemble achieves an average macro F1 of \textbf{0.9280} across four hospital centres, outperforming the best single model (F1$=$0.8969) by +0.031, demonstrating that heterogeneous architectures combined with source-aware calibration are essential for robust multi-site medical image classification.
\end{abstract}    
\section{Introduction}
With the rapid worldwide spread of COVID-19, the medical community was shown to have critical limitations in current diagnostic workflows and called for an urgent need to conduct research to develop automated screening tools~\cite{kollias2021mia}. Although Reverse-Transcriptase Polymerase Chain Reaction (RT-PCR) tests are the gold standard for COVID-19 diagnosis, they take days to produce results and are far too often negative when a patient actually has COVID-19~\cite{kollias2023ai}, therefore resulting in unnecessary transmission of the disease or negative or delayed treatment of infected patients. Computed Tomography (CT) is another form of COVID-19 diagnostic tool that is complementary to RT-PCR and can quickly and accurately identify patient lung pathology associated with COVID-19 using ground-glass opacities (GGO), consolidations with an associated air bronchogram, and/or ``crazy paving'' or fine, wavy, or irregular septations~\cite{kollias2021mia, kollias2023ai}.

Deep learning has shown tremendous potential in automating the process of detecting COVID-19 from CT scans, with early studies conducted by Kollias~\etal~\cite{kollias2018deep} establishing the architecture of deep neural networks for use in the prediction of healthcare outcomes and later research exploring ever more sophisticated methods of using 3D chest CT scans for the diagnosis of COVID-19~\cite{kollias2020deep, kollias2020transparent, kollias2021mia}. More recently, research has been completed regarding the use of deep learning vision-language models for CT segmentation~\cite{kollias2024sam2clip2sam}, the application of domain adaptation with a focus on fairness issues~\cite{kollias2024domain}, and the Pharos-AFE-AIMI challenge which addresses multi-source generalization~\cite{kollias2025pharos}.

A fundamental challenge in CT-based diagnosis is that each scan is a 3D volume comprising tens to hundreds of slices, yet only a subset of these slices contains diagnostically relevant patterns such as ground-glass opacities or consolidations. In clinical practice, only scan-level labels (COVID-positive or negative) are available, as requiring radiologists to annotate individual slices is prohibitively expensive and time-consuming. Multiple Instance Learning (MIL)~\cite{ilse2018attention} provides a natural framework for this weakly supervised setting: the scan is treated as a ``bag'' of slice ``instances,'' and the model learns to classify the entire bag without requiring instance-level annotations. Attention-based MIL is particularly well-suited to this task, as the learned attention weights reveal which slices the model considers most informative, providing interpretability that is critical for clinical trust. By automatically identifying and weighting the most discriminative slices, attention-based MIL enables efficient scan-level classification while preserving the ability to localize pathological regions within the volume.

One of the challenges of model generalization is deploying them across many different data sources with distinct differences between hospitals, including scanners and protocol types as well as the makeup of the patient population. There is no one model architecture or pretraining method that has been observed to generalize across all data sources. Instead, we find robust performance results from combining many different models, which contribute different inductive biases and representational strengths. In this research effort, we are proposing a three-pronged approach to addressing this challenge:

\begin{enumerate}[leftmargin=*,itemsep=2pt]
    \item We built a \textbf{collection of nine heterogeneous models}, which consists of three inference paradigms: self-supervised DINOv2~\cite{oquab2024dinov2} with slice-level aggregations, RadImageNet-pretrained DenseNet-121~\cite{huang2017densenet, mei2022radimagenet} with slice-level sigmoid aggregations, and Gated Attention MIL~\cite{ilse2018attention} with three different CNN backbone families. These maximize both architectural diversity as well as representational diversity.
    \item We developed a \textbf{regularization methodology} that resulted in the ratio of the validation/training loss falling from 35$\times$ to less than 3$\times$, allowing us to train for many epochs beyond only 1--2 epochs.
    \item We calibrated the \textbf{threshold for decision-making on a source-by-source basis}, which optimizes each hospital's output distribution thresholds and demonstrates a +0.14 F1 score improvement over global thresholding.
\end{enumerate}
The ensemble design itself is central to these gains: combining architectures with distinct representational strengths proves more robust across hospital centers than scaling any single model family.

Overall these results represent an average F1 value of 0.9280 averaged over all four hospital locations; this exceeds the best single system (DINOv2, F1=0.8969) by 0.031 points. The largest contribution to performance improvement was through balancing the thresholds used at each of the sources which accounted for a +0.14 increase in F1 compared to a shared global threshold; therefore, post process technique must account for the different sources of CT data when evaluating an ensemble of models using an average across differing data sources.

The remainder of this paper is structured as follows: Section 2 covers related work, Section 3 presents the dataset we used, Section 4 explains the methods, Section 5 describes the results, Section 6 addresses concerns and strengths, and Section 7 concludes.
\section{Related Works}

COVID-19 chest CT became a major test case for deep learning very early in the pandemic. Early papers showed that CT scans could be used not only for COVID-19 detection but also, in some settings, for severity estimation~\cite{kollias2021mia, arsenos2022large, kollias2022ai, kollias2023ai}. That mattered because it gave the field a concrete imaging problem with usable benchmarks and enough signal for automated prediction~\cite{kollias2021mia, arsenos2022large}.

Scan-level labeling creates another issue. In many CT datasets, the label is attached to the whole scan even though only part of the volume may contain the most useful evidence. This is where Multiple Instance Learning becomes relevant. Instead of forcing every slice to contribute equally, MIL allows the model to learn which slices matter more when making the final scan-level decision~\cite{ilse2018attention}. Attention-based MIL is especially relevant for this reason~\cite{ilse2018attention}.

Transfer learning is also closely tied to this literature. Medical datasets are expensive to label, so researchers have looked beyond fully supervised training for better visual features. Self-supervised learning is one answer~\cite{oquab2024dinov2}. Radiology-specific pretraining is another~\cite{mei2022radimagenet}. Both try to improve feature quality in settings where annotation is limited and data from different hospitals do not look exactly the same.

So the problem is broader now than early benchmark classification. COVID-19 CT work sits alongside questions about weak supervision, transfer across datasets, and whether a model trained in one setting still holds up in another~\cite{kollias2024domain, gerogiannis2024covid, kollias2025pharos}.

\section{Dataset}
\label{sec:Dataset}

The dataset used for this research is part of the “Multi-Source COVID-19 Detection Challenge”, organized by PHAROS AI Factory for Medical Imaging \& Healthcare (PHAROS-AIF-MIH) in conjunction with the IEEE Computer Vision and Pattern Recognition Conference (CVPR) 2026.
The dataset consists of 3D chest CT scans from four hospital data centers with distinct scanner hardware and acquisition protocols. Each scan is stored as a folder of JPEG slices (typically 50–700 per scan at 512$\times$512 resolution), split into ~1,222 training and 308 validation scans. In addition to image folders, the organizers provide CSV metadata files. These indicate the COVID label and the source identifier in {0, 1, 2, 3} for each scan in the dataset.

\begin{figure}[t]
  \centering
  \begin{subfigure}{0.32\linewidth}
    \centering
    \includegraphics[width=\linewidth]{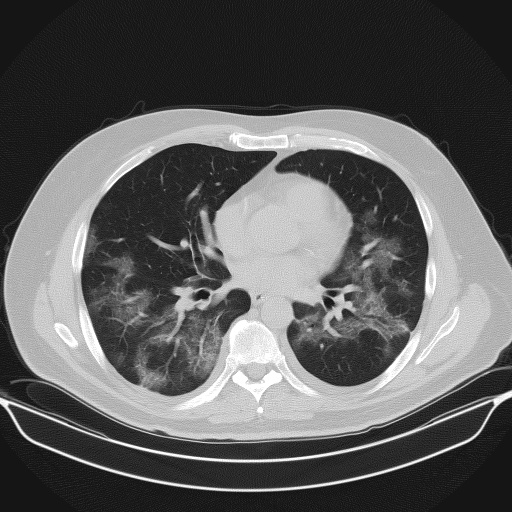}
  \end{subfigure}
  \hfill
  \begin{subfigure}{0.32\linewidth}
    \centering
    \includegraphics[width=\linewidth]{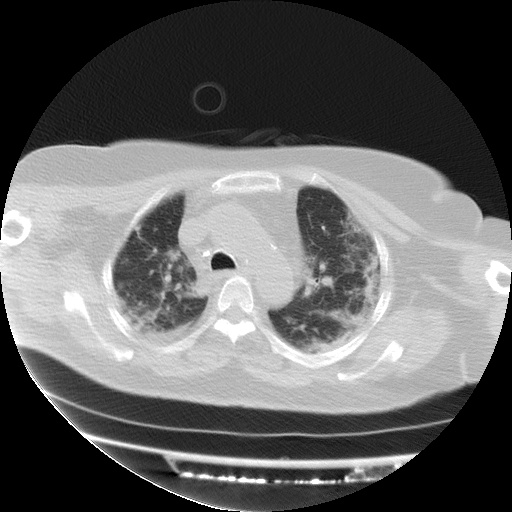}
  \end{subfigure}
  \hfill
  \begin{subfigure}{0.32\linewidth}
    \centering
    \includegraphics[width=\linewidth]{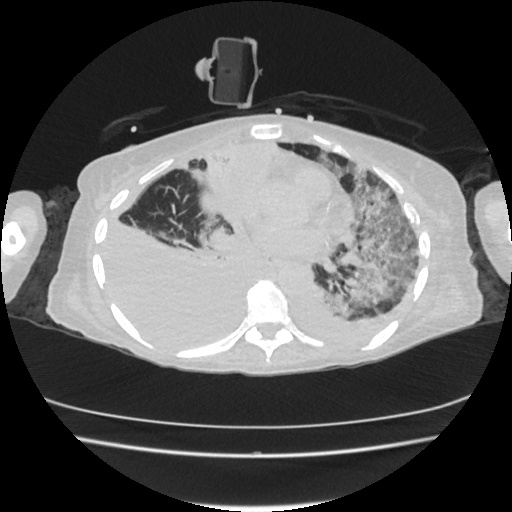}
  \end{subfigure}
  \caption{Representative COVID-positive CT slices from three different data centers in the validation set, each showing ground-glass opacities and multi-focal involvement with differing scanner characteristics and acquisition protocols.}
  \label{fig:covid_examples_centres}
\end{figure}

This research followed the official training and evaluation split of the Multi-Source COVID-19 Detection Challenge. This split yields a total of 1,224 training scans (564 COVID, 660 non-COVID) and $308$ validation scans (128 COVID, 180 non-COVID) as referenced in the table below. In particular, COVID cases from source 2 are rare (39 training, 0 validation scans). In addition to the training and validation datasets, the organizers also provide an unlabeled test dataset to report the final inference of the model in the form of a CSV file. Center 2's limited observations in the training and validation sets created a challenging low-resource constraint that strongly influenced our per-center macro F1. It also motivated the center-balanced sampling strategy that we describe in the methodology.

\begin{table}[ht]
  \centering
  \caption{\textbf{Number of CT scans per source and split in the PHAROS multi-source COVID-19 dataset.}}
  \label{tab:dataset_counts}
  \small
  \setlength{\tabcolsep}{4pt}
  \begin{tabular}{@{}lcccc@{}}
    \toprule
    \textbf{Source} &
    \shortstack{\textbf{Train}\\\textbf{COVID}} &
    \shortstack{\textbf{Train}\\\textbf{non-COVID}} &
    \shortstack{\textbf{Validation}\\\textbf{COVID}} &
    \shortstack{\textbf{Validation}\\\textbf{non-COVID}} \\
    \midrule
    0 & 175 & 165 & 43 & 45 \\
    1 & 175 & 165 & 43 & 45 \\
    2 & 39  & 165 & 0  & 45 \\
    3 & 175 & 165 & 42 & 45 \\
    \midrule
    \textbf{Total} & \textbf{564} & \textbf{660} & \textbf{128} & \textbf{180} \\
    \bottomrule
  \end{tabular}
\end{table}

The evaluation metric used to assess the models is as follows:

\begin{equation}
  P = \frac{1}{4} \sum_{i=0}^{3} \frac{F1^{i}_{\mathrm{covid}} + F1^{i}_{\mathrm{noncovid}}}{2}.
  \label{eq:also-important}
\end{equation}

This criteria was provided by PHAROS AI Factory for Medical Imaging \& Healthcare (PHAROS-AIF-MIH) as part of the “Multi-Source COVID-19 Detection Challenge” ~\cite{kollias2025pharos}. It yields a fair way to aggregate a cumulative F-1 score by center and label.

\section{Methods}

\subsection{Overview}

We present a heterogeneous ensemble of nine models designed to capture complementary information at both the slice and scan levels. The ensemble combines three inference paradigms: (1) a self-supervised DINOv2 Vision Transformer with slice-level sigmoid aggregation, (2) a RadImageNet-pretrained DenseNet-121 with slice-level sigmoid aggregation, and (3) seven Gated Attention Multiple Instance Learning (MIL) models with scan-level softmax classification. The MIL branch includes EfficientNet-B3~\cite{tan2019efficientnet}, ConvNeXt-Tiny~\cite{liu2022convnext}, and EfficientNetV2-S~\cite{tan2021efficientnetv2} backbones. Final predictions are obtained by combining model probabilities through ensemble fusion and validation-time threshold calibration.

This design was motivated by diversity at four levels: backbone architecture, pretraining source, inference paradigm, and random seed. DINOv2 contributes global self-attention and self-supervised features~\cite{oquab2024dinov2}; DenseNet contributes dense convolutional feature reuse and radiology-domain pretraining~\cite{huang2017densenet, mei2022radimagenet}; the MIL branch contributes learned scan-level weighting over slices via gated attention~\cite{ilse2018attention}. The final ensemble therefore combines models that make predictions in different ways rather than relying on repeated instances of the same architecture.

\subsection{Preprocessing and Augmentation}

All input slices are first resized to $256\times256$ pixels. This gives a common canvas across models and allows spatial augmentation before the final task-specific crop. For the DINOv2 and DenseNet branches, a random crop is applied during training or a center crop during validation to obtain $224\times224$ inputs. This is preferable to starting directly at $224\times224$, because the larger intermediate resize preserves more field of view and makes the crop itself part of the augmentation pipeline. For the slice-level branches, this also keeps the pre-processing consistent across scans with different native in-plane extents before the final crop is sampled. Gated Attention MIL models operate on full $256\times256$ resolution to preserve all evidence and let attention softly weight it.

All models are normalized using standard ImageNet statistics, using mean $[0.485, 0.456, 0.406]$ and standard deviation $[0.229, 0.224, 0.225]$, following the preprocessing convention used by the pretrained backbones employed in~\cite{dosovitskiy2021vit, mei2022radimagenet, wightman2021timm}.

During training, we use a compact augmentation pipeline that preserves radiological structure while improving robustness to scanner and protocol variation. The pipeline includes horizontal flipping with probability $0.5$, in-plane rotation by up to $\pm 15^\circ$ with probability $0.5$, random brightness and contrast adjustment in the range $\pm 0.2$ with probability $0.5$, and Gaussian blur with kernel size 3--7 and probability $0.1$. For the DINOv2 and DenseNet branches, cropping is applied after augmentation so that rotations do not introduce black-border artifacts into the final image. This design keeps the transformations simple and anatomically plausible while still exposing the models to moderate variation in orientation and appearance.

\subsection{Center-Stratified Sampling Per Epoch}
\label{subsec:center_stratified}

For the slice-level branches, training is performed on individual slices rather than whole scans. Each training scan is first converted into a set of slice samples, and when a scan contains more than the configured maximum number of slices, we retain a uniformly spaced subset along the axial axis. This preserves coverage across the full superior--inferior extent of the volume while preventing very long scans from dominating the slice pool.

To reduce source imbalance, the training loader uses a center-stratified batch sampler implemented over slice indices. At the start of each epoch, slices are grouped by medical center, each center-specific pool is shuffled, and smaller centers are oversampled with replacement until they match the size of the largest center. Mini-batches are then formed by drawing approximately equal numbers of slices from each center and shuffling them within the batch. As a result, every epoch exposes the model to all centers in a balanced way rather than allowing the largest source to dominate optimization.

This sampler balances batches by source rather than by class. Class proportions therefore remain tied to the underlying distribution within each center, while inter-center imbalance is corrected explicitly. In our setting, this is important because the challenge metric averages performance across centers, so optimization should not be driven disproportionately by the most common source. One difference in the DenseNet-121 branch is that the sampling is center and epoch based to ensure each epoch has an equal representation of positive and negative cases.

\subsection{Slice-Averaging Branches}

The first two branches share the same inference logic: they process slices independently, convert slice logits to probabilities with the sigmoid function $\sigma(\cdot)$, and average these probabilities across the full scan. They differ mainly in backbone architecture and pretraining source.

\paragraph{DINOv2 Vision Transformer.}
The first slice-averaging branch uses DINOv2 ViT-B/14~\cite{oquab2024dinov2}, a self-supervised Vision Transformer with 86M parameters, a 14$\times$14 patch size, and a 768-dimensional CLS embedding. We replace the original head with a binary classifier of the form
\[
\mathrm{Dropout}(0.4)\rightarrow \mathrm{Linear}(768,1),
\]
which produces one logit per slice. Training is carried out in two phases. In Phase~1, the backbone is frozen and only the classifier head is trained. In Phase~2, we progressively unfreeze the last two and then the last four transformer blocks using discriminative learning rates. Both phases use AdamW~\cite{loshchilov2019adamw} with cosine scheduling. For both phases, the slice-level training batches follow the center-stratified sampling procedure described in \cref{subsec:center_stratified}.

\paragraph{DenseNet-121 with RadImageNet.}
The second slice-averaging branch uses DenseNet-121~\cite{huang2017densenet} initialized from RadImageNet~\cite{mei2022radimagenet}. The final DenseNet representation is 1024-dimensional and is passed through a binary classifier
\[
\mathrm{Dropout}(0.4)\rightarrow \mathrm{Linear}(1024,1).
\]
As with DINOv2, training proceeds in two stages: head-only fine-tuning followed by progressive unfreezing of deeper blocks. In this case, Phase~2 unfreezes \texttt{denseblock4} and later \texttt{denseblock3} with smaller backbone learning rates than the classification head. DenseNet training uses the same center-stratified slice sampling as the DINOv2 branch.

For both slice-averaging branches, all slices in a scan are processed independently and converted to sigmoid probabilities. The scan-level COVID score is the arithmetic mean across slices:
\begin{equation}
P_{\mathrm{covid}}=\frac{1}{N}\sum_{i=1}^{N}\sigma(f(x_i)),
\label{eq:slice_avg}
\end{equation}
where $x_i$ is the $i$-th slice, $f(\cdot)$ denotes the corresponding slice classifier, $N$ is the number of slices in the scan, and $\sigma(\cdot)$ is the sigmoid function.

The two branches differ in what $f(\cdot)$ represents. For DINOv2, $f(\cdot)$ is the ViT-B/14 backbone with a 768-to-1 head; for DenseNet, it is the RadImageNet-pretrained DenseNet-121 backbone with a 1024-to-1 head. Both branches use four-view test-time augmentation (TTA). The views are the original slice, a horizontal flip, a $+15^\circ$ rotation, and a $-15^\circ$ rotation. These two branches contribute a complementary perspective to the MIL models: they do not learn explicit attention over slices, but instead rely on strong slice-level discriminative features and a simple, stable scan-level aggregation rule.

\subsection{Gated Attention MIL Branch}

The remaining seven models use a two-phase Gated Attention MIL pipeline. Unlike the slice-averaging branches, these models produce one scan-level prediction directly from a set of slice embeddings. We use EfficientNet-B3, ConvNeXt-Tiny, and EfficientNetV2-S backbones, which yield 1536-, 768-, and 1280-dimensional embeddings, respectively. The seven models comprise three EfficientNet-B3 seed variants plus one EfficientNet-B3 SWA variant, one ConvNeXt-Tiny model, and one EfficientNetV2-S model plus one EfficientNetV2-S SWA variant.

\paragraph{Phase 1: slice-level pretraining.}
Before training the full MIL model, each backbone is first pretrained at the slice level. A lightweight slice classifier is attached to the backbone and trained to distinguish COVID from non-COVID slices. For EfficientNet-B3, for example, this head is
\[
\mathrm{Dropout}(0.3)\rightarrow \mathrm{Linear}(1536,2).
\]
This phase runs for 20 epochs using AdamW with learning rate $10^{-4}$, weight decay $0.01$, cosine decay with a 2-epoch warmup, cross-entropy loss with label smoothing $\varepsilon=0.1$, and center-stratified balanced slice sampling during pretraining. The purpose of this stage is to learn clinically meaningful slice representations before introducing scan-level attention.

\paragraph{Phase 2: scan-level MIL training.}
In the second phase, each scan is represented as a set of $K$ slices. During training, the MIL branch uses $K=24$ slices per scan, while the final evaluation setting uses $K=48$ slices per scan. Given slice embeddings $\mathbf{h}_1,\ldots,\mathbf{h}_K$, gated attention pooling~\cite{ilse2018attention} computes a learned weight for each slice:
\begin{align}
\mathbf{v}_k &= \tanh(\mathbf{W}_V\mathbf{h}_k), \\
\mathbf{u}_k &= \sigma(\mathbf{W}_U\mathbf{h}_k), \\
a_k &= \frac{\exp\!\left(\mathbf{w}^{\top}(\mathbf{v}_k\odot \mathbf{u}_k)\right)}
{\sum_{j=1}^{K}\exp\!\left(\mathbf{w}^{\top}(\mathbf{v}_j\odot \mathbf{u}_j)\right)},
\label{eq:gated_attention}
\end{align}
where $\odot$ denotes element-wise multiplication. The scan embedding is then
\begin{equation}
\mathbf{z}=\sum_{k=1}^{K} a_k\mathbf{h}_k.
\label{eq:scan_embedding}
\end{equation}

The pooled representation is passed to a two-layer classifier of the form
\[
\mathrm{Linear}(d,h)\rightarrow \mathrm{ReLU}\rightarrow \mathrm{Dropout}(0.5)\rightarrow \mathrm{Linear}(h,2),
\]
where $d$ is the backbone embedding dimension and $h$ is a hidden dimension. We use $h=512$ for EfficientNet-B3 and EfficientNetV2-S, and $h=384$ for ConvNeXt-Tiny.

The MIL models are trained for up to 30 epochs with early stopping patience 8, mixed-precision FP16 training, and a short freeze period for the backbone during the first 3 epochs. AdamW is used with differential learning rates: $10^{-6}$ for the backbone and $10^{-5}$ for the attention and classifier layers, with weight decay $0.05$. The effective batch size is 32, implemented through a physical batch size of 2 and gradient accumulation over 16 steps. Because scan-level training is memory intensive, we use gradient checkpointing and chunked forward passes through the backbone. In our implementation, slice batches are processed in chunks of 8, which substantially reduces GPU memory usage without changing the model formulation.

For test-time augmentation (TTA), the MIL models use four-view flip-based TTA: original, horizontal flip, vertical flip, and a combined horizontal-plus-vertical flip. Logits are averaged across these views before the final softmax is applied.

\subsection{Loss Functions and Regularization}

Several regularization choices were used to stabilize training across the ensemble.

\paragraph{Focal Loss.}
For the MIL branch, Focal Loss~\cite{lin2017focal} replaces standard cross-entropy during Phase~2 in order to emphasize difficult examples. This is useful in our setting because easy scans quickly dominate the gradient under standard cross-entropy, whereas the challenge is largely determined by harder borderline cases and centre-specific shifts. Its form is
\begin{equation}
\mathrm{FL}(p_t)=-\alpha_t(1-p_t)^{\gamma}\log(p_t),
\label{eq:focal_loss}
\end{equation}
with $\gamma=2.0$ and $\alpha=[0.55,0.45]$, mildly up-weighting the COVID class.

\paragraph{Embedding-level Mixup.}
Mixup~\cite{zhang2018mixup} is applied after attention pooling rather than on raw slices. If two scan embeddings are $\mathbf{z}_a$ and $\mathbf{z}_b$, then the mixed embedding is
\begin{equation}
\mathbf{z}_{\mathrm{mix}}=\lambda \mathbf{z}_a + (1-\lambda)\mathbf{z}_b,
\label{eq:mixup}
\end{equation}
where $\lambda\sim\mathrm{Beta}(0.2,0.2)$. We use embedding-level rather than image-level Mixup because direct interpolation between CT slices can create anatomically implausible inputs, whereas interpolation in the learned embedding space acts as a smoother regularizer.

\paragraph{Additional regularization.}
We further use dropout, label smoothing, and stochastic depth (drop path rate 0.3)~\cite{huang2016stochastic}. For selected models, we apply Stochastic Weight Averaging (SWA)~\cite{izmailov2018swa} for five additional epochs after the main training phase in order to favor broader optima and improve cross-center generalization.

\subsection{Ensemble Construction and Model Selection}

The final ensemble contains nine models. For clarity, we summarize them below:
\begin{itemize}[leftmargin=*,itemsep=2pt]
    \item 1 DINOv2 slice-level model;
    \item 1 RadImageNet DenseNet-121 slice-level model;
    \item 4 EfficientNet-B3 Gated Attention MIL models;
    \item 1 ConvNeXt-Tiny Gated Attention MIL model;
    \item 2 EfficientNetV2-S Gated Attention MIL models.
\end{itemize}

The backbone diversity is intentional. DINOv2 and DenseNet inject complementary pretrained representations, while the MIL branch provides learned scan-level weighting and multiple local minima through seed and SWA variants. Before fusion, all outputs are converted to a common probability space. The DINOv2 and DenseNet models already yield scalar scan-level probabilities. The MIL models produce 2-class softmax outputs, from which we take the COVID probability.

Our final fusion rule is score-weighted probability averaging. If model $m$ produces a probability $p_m$ and has validation macro F1 score $s_m$, then the ensemble probability is
\begin{equation}
p_{\mathrm{ens}}=\sum_{m=1}^{M}\frac{s_m}{\sum_{j=1}^{M}s_j}\,p_m,
\label{eq:ensemble-weighted}
\end{equation}
where $M=9$ in the final system. This gives higher-performing constituent models greater influence while retaining contributions from diverse model families.

\paragraph{Compute infrastructure.}
All experiments were conducted on an HPC cluster. The SLURM jobs in our implementation request one NVIDIA RTX A6000 GPU per job, 64~GB of host memory, and 8 CPU cores for validation and test ensemble runs. The ensemble implementation launches one worker process per model and supports round-robin assignment across the set of visible GPUs, so the same code can exploit parallel multi-GPU execution when more than one device is made available.

\subsection{Threshold Calibration and Alternative Ensemble Strategies}

The PHAROS challenge metric averages macro F1 across sources, so calibration matters as much as raw probability ranking. We therefore tune thresholds on the validation set rather than fixing them at 0.5. In the final system, threshold selection is performed separately for each source. For source $s$, we sweep $t_s \in [0.20, 0.80]$ in steps of 0.005 and select the threshold that maximizes macro F1 on that source's validation scans. This source-aware calibration compensates for systematic shifts in score distributions across hospitals.

We evaluated multiple ensemble strategies during validation:
\begin{itemize}[leftmargin=*,itemsep=2pt]
    \item \textbf{Uniform probability averaging}: all model probabilities receive equal weight.
    \item \textbf{Score-weighted probability averaging}: probabilities are combined according to validation F1, as in \cref{eq:ensemble-weighted}.
    \item \textbf{Majority vote}: each model casts a hard decision after thresholding, and the final prediction is the majority class.
    \item \textbf{Global thresholding versus per-source thresholding}: we compared a single validation-tuned threshold against source-specific thresholds.
\end{itemize}

These comparisons were carried out on the validation set using the same challenge metric used for the final submission. This procedure allowed us to compare not only individual models but also ensemble fusion rules under the exact evaluation criteria of the competition. The final chosen system combined score-weighted probability averaging with per-source threshold calibration, yielding the best validation F1 score of 0.9280.
\section{Results}
\label{sec:Results}

\begin{table}[t]
\centering
\caption{Best validation F1 scores by pre-trained model family.}
\label{tab:individual}
\begin{tabular}{lc}
\toprule
\textbf{Backbone} & \textbf{F1} \\
\midrule
EffNet-B3       & 0.856 \\
DenseNet-121    & 0.865 \\
ConvNeXt-T      & 0.869 \\
EffNetV2-S      & 0.877 \\
DINOv2 ViT-B/14 & 0.910 \\
\textbf{Ensemble Model}  & \textbf{0.928} \\
\bottomrule
\end{tabular}
\end{table}

\begin{figure}[t]
\centering
\includegraphics[width=0.64\columnwidth]{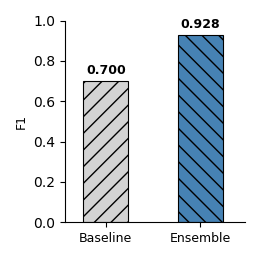}
\caption{Ensemble approach vs.\ baseline ResNet50 on a similar dataset.}
\label{fig:training_samples}
\end{figure}

This section evaluates the final nine-model ensemble with the validation split. This validation split has ground-truth labels that are necessary to calculate the final challenge metric. The reported score (0.928) is the source-wise macro F1 averaged across the four data centers. As labels for the test split are not available, test-time experiments are limited to inference and probability generation (no final F1 score can be calculated).

We conduct validation experiments using the ensemble configuration described in the methodology, including one DINOv2 slice-level model, one RadImageNet-pretrained DenseNet-121 slice-level model, four EfficientNet-B3 Gated Attention MIL models, one ConvNeXt-Tiny Gated Attention MIL model, and two EfficientNetV2-S Gated Attention MIL models. Each model produces a scan-level COVID probability, after which ensemble predictions are formed via probability averaging. To achieve an F1 score of 0.928 with the validation split, we also applied threshold calibration. The direct comparison of the ensemble model with the individual constituent models is shown in Table 2. On the other hand, figure 2 provides a contextual comparison with the 2021 benchmark model reported by Kollias \etal~\cite{kollias2021mia}. Finally, figure 3 depicts the progression of F1 scores over the different training phases. Here, for DinoV2 and DenseNet-121, phase 1 refers to the head training phase with a frozen backbone, and for the other models, phase 1 refers to the slice-level pre-train. Phase 2a is the unfreezing of the two transformer blocks and the last dense blocks in DinoV2 and DenseNet-121 respectively. However, in the other MIL-based models, phase 2 is the end-to-end MIL aggregation. Phase 2b is only for DinoV2 and DenseNet-121, in which we unfreeze more blocks in the backbone. The 'Final' label on the X-axis is the final reported F1 after the evaluation on the validation dataset. This figure does not include the ensemble model, as it was simply a collation of trained models. It did not require phase-by-phase training and only required a final evaluation at the end. Together, these validation experiments show that the ensemble achieves the strongest overall performance among the tested configurations, supporting the use of complementary architectures rather than the dependence on a single backbone family.

\begin{figure}[t]
\centering
\includegraphics[width=\columnwidth]{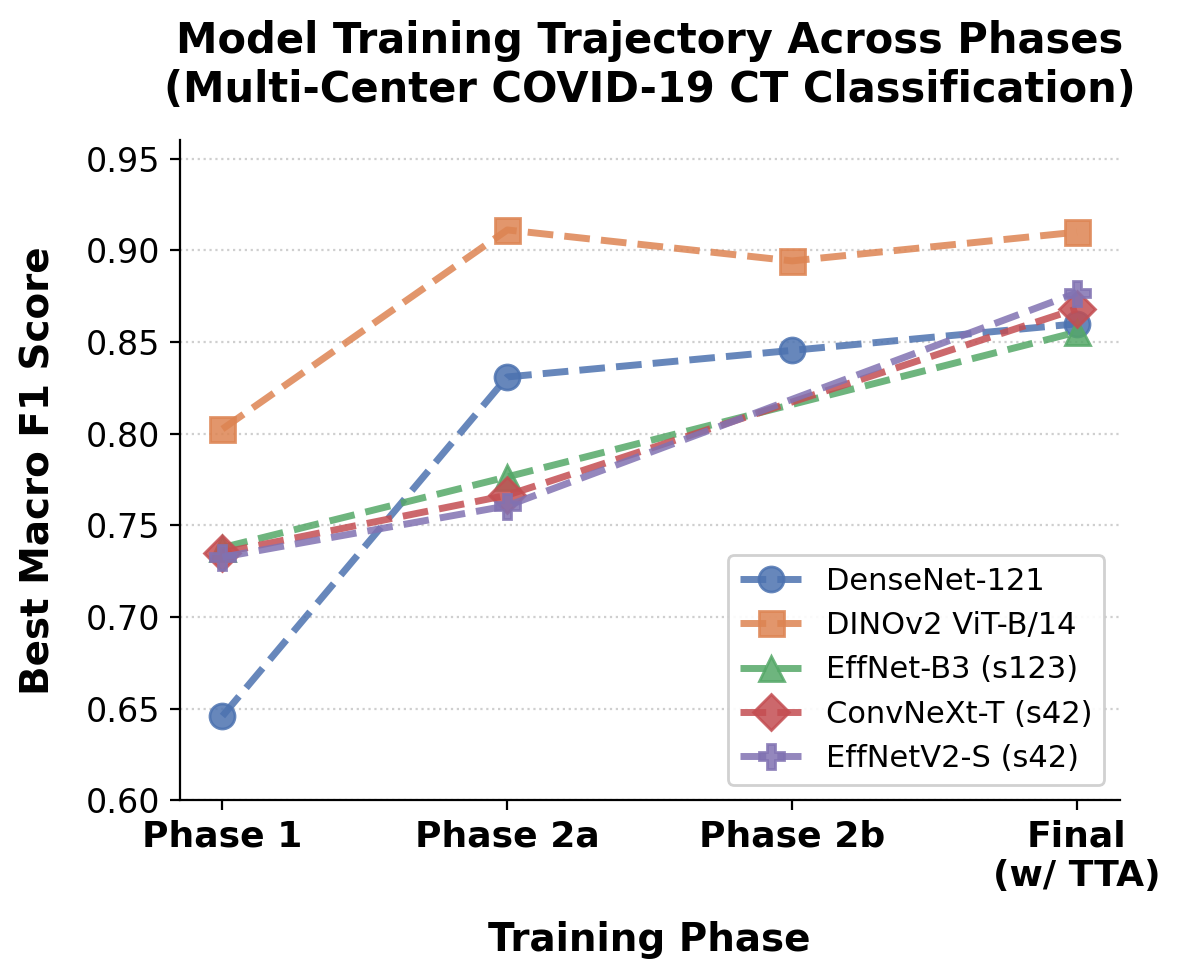}
\caption{Phase-by-phase F1 progression of different models}
\label{fig:success}
\end{figure}

\begin{figure}[t]
\centering
\includegraphics[width=\columnwidth]{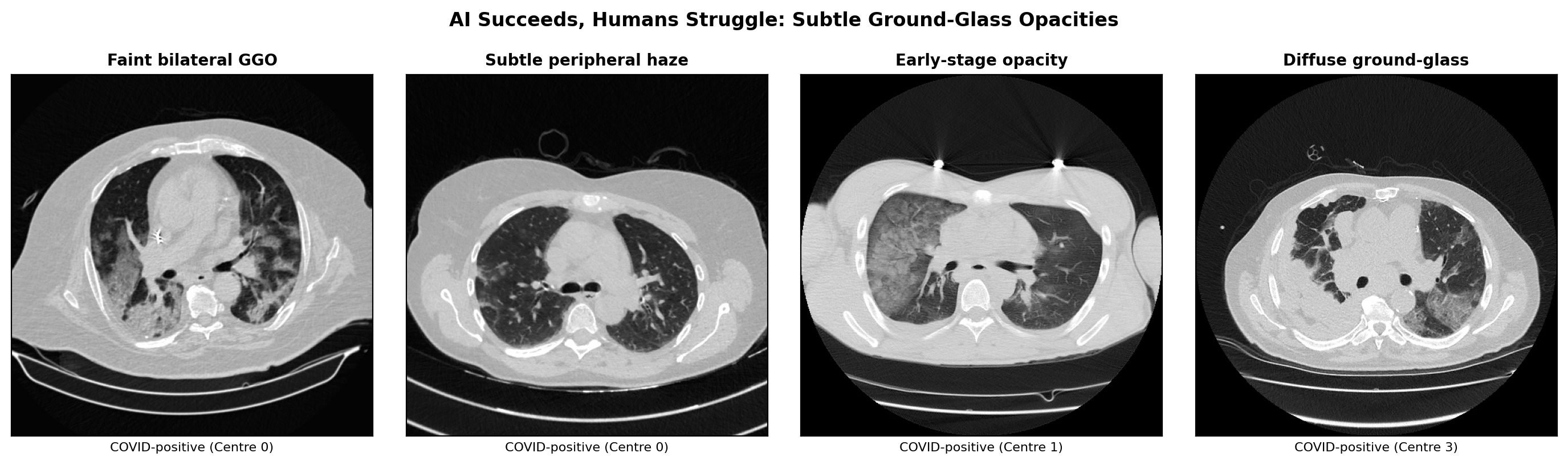}
\caption{COVID-positive CT slices with subtle ground-glass opacities (GGO). These faint, poorly-contrasted regions are easily overlooked during rapid human screening but are consistently detected by the attention mechanism across multiple slices.}
\label{fig:success}
\end{figure}

\begin{figure}[t]
\centering
\includegraphics[width=\columnwidth]{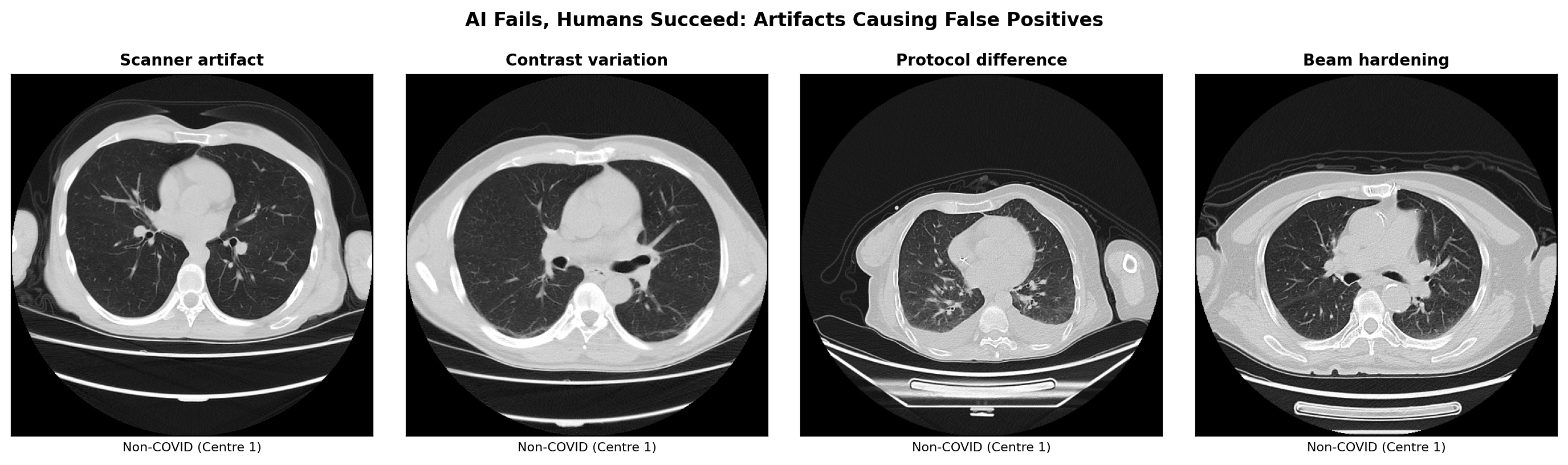}
\caption{Non-COVID CT slices from Centre~1 that the model tends to misclassify as COVID-positive. Scanner-specific artifacts, contrast variations, and protocol differences create patterns that confuse the model but are easily recognized as non-pathological by trained radiologists.}
\label{fig:failure}
\end{figure}

\section{Discussion}
\label{sec:Discussion}

An important limitation of the data is its small size and the imbalance across centers, particularly the low number of COVID-positive cases in source 2. This makes both training and validation more challenging and can increase performance variability across centers. Despite these constraints, the ensemble achieves strong overall performance, suggesting that combining complementary pretrained and fine-tuned models from different families improves robustness relative to any single constituent model alone. At the same time, comparisons with earlier external benchmarks should be interpreted cautiously whenever dataset composition, pre-processing, or evaluation protocols differ.

As visible in Figure 4, there are cases in which the model performs exceptionally well even when the finding is difficult for a human to spot. As reasoned before, we believe this is due to our meticulous approach combining strategies like center-stratified slice sampling, augmentations of data, progressive freezing and unfreezing of blocks, and gated attention MIL. This ensures that several different models trained in different ways pick on unique characteristics that contribute to the model's overall performance.

However, there are also cases, such as those shown in Figure 5, where our model makes an incorrect prediction despite patterns that are clearly indicative of the opposite label. Some of these discrepancies can be attributed to the inherent nuance introduced by a multi-source dataset, including scanner-specific artifacts, contrast variations, and protocol differences that may confuse the model. One way to mitigate this issue could be to incorporate a CLIP-like image segmentation tool into our proposed framework to help the model focus more effectively on the most important portions of the image~\cite{zhao2023clipsurvey}.

In terms of practical usage, all experiments were conducted on an HPC cluster. The checked SLURM jobs in our implementation request one NVIDIA RTX A6000 GPU per job, while the ensemble code supports parallel multi-GPU inference by assigning separate worker processes to available GPUs in round-robin fashion. This makes the pipeline practical for shared cluster environments and feasible for deployment settings where several open-source constituent models can be evaluated in parallel to produce timely scan-level predictions.

\section{Conclusion}
We presented a nine-model ensemble for multi-source COVID-19 detection from 3D chest CT scans, which combines self-supervised DINOv2, RadImageNet-pretrained DenseNet-121, and also seven Gated Attention MIL models over three CNN backbone families. This ensemble capitalizes on the diversity of four axes: architecture, pretraining strategy, inference paradigm, and training seed. This allowed the model to achieve a \textbf{0.9280} average macro F1 score across all four hospital centres. This ended up outperforming the best individual model (DINOv2, F1$=$0.8969) by +0.031. This result was achieved due to three technical developments: (1) the heterogeneous ensemble, which combines models that have different types of inductive bias (e.g., global self-attention, reuse of dense convolutional features, and learned attention pooling), so that the failure mode of any one single model does not dominate; (2) the use of a regularisation technique that includes Focal Loss, embedding based Mixup and domain-aware augmentations reducing the validation to training loss ratio from 35$\times$ to less than 3$\times$; and (3) Source Specific Threshold Calibration, which results in a +0.14 F1 improvement compared to using one global threshold for all sources.

Our results demonstrate that no single architecture dominates across all hospital centres. Instead, it combines models with distinct inductive biases. The global self-attention (DINOv2), dense feature reuse (DenseNet), and learned attention pooling (MIL) provide the strongest multi-source generalization. Future work can address the remaining failure modes: motion artifacts and beam hardening through artifact-aware preprocessing, 3D convolutional/transformer architectures for richer volumetric context, and federated learning for privacy-preserving multi-centre training~\cite{kollias2024domain, kollias2025pharos}.

{
    \small
    \bibliographystyle{ieeenat_fullname}
    \bibliography{main}
}


\end{document}